\documentstyle [12pt]{article}
\begin{document}
~~~~~~~~~~~~~~~~~~~~~~~~~~~~~~~~~~~~~~~~~~~~~~~~~~~~~~~~~~~~~~~~
~~~~~~~~~~~~~~~~~~~~~~~~~~May 1996
\vskip 1.8cm
\centerline {\bf SUPERLUMINAL MATTER AND HIGH-ENERGY COSMIC RAYS}
\vskip 1.5cm
\centerline {\bf L. GONZALEZ-MESTRES}
\vskip 4mm
\centerline {Laboratoire de Physique Corpusculaire, 
Coll\`ege de France,}
\centerline {11 pl. Marcellin-Berthelot, 75231 Paris Cedex 05 , 
France}
\vskip 1.5cm
High-energy cosmic ray events present important challenges to particle 
astrophysics. Their nature and origin are often
not well understood and, as they occur in an energy domain not 
accessible to particle accelerators, there is no
clear guide to their interpretation.
We recently proposed [1] a 
new, non-tachyonic, possible class of superluminal particles which 
may play an important cosmological role
[2]. In this note, we would like to point out that such particles, 
if they exist, are able to produce very high energy
cosmic rays compatible with unexplained discoveries reported in 
the literature.
\vskip 6mm
{\bf 1. THE NEW SUPERLUMINAL PARTICLES}
\vskip 5mm
Since we know from particle physics that vacuum is not an empty entity 
but a material medium with many internal degrees of
freedom, there is no fundamental reason to interpret Lorentz 
invariance as an absolute property of space and time
previous to the existence of matter. 
Instead, we may consider it as an approximate dynamical property of 
the equations describing
an important sector of matter (the "ordinary" sector, i.e. the 
particles whose critical speed in vacuum is the speed of light)
above a certain distance scale ($k_0^{-1}$ , where $k_0$ is a 
critical wave vector scale) related to
a fundamental phase transition (at $T \approx k^{-1} \hbar c k_0$ , 
$k$ = Boltzmann constant, $\hbar $ = Plack constant [2]).
Then, ordinary Lorentz 
invariance would not necessarily be universal and other degrees of 
freedom can exist leading to particles with
different critical speeds in vacuum (the "superluminal" sectors, 
with critical speeds $c_i \gg c$ where the subscript $i$
stands for the $i$-th superluminal sector).
If $m$ is the inertial mass, a particle belonging to the $i$-th 
superluminal sector would have [1] rest energy:
\equation
E_{rest}~=~m~c_i^2
\endequation
leading potentially to very large rest energies. 
Annihilation of pairs of superluminal particles into ordinary ones can
therefore release very large kinetic energies and provide a new 
source of high-energy cosmic rays.
If superluminal matter exists, it 
is likely to have dominated the evolution of the very early 
Universe [2] and be still present
playing a crucial role at large scales.
The study of fundamental cosmological problems (e.g. nature of the big 
bang, horizon problem, cosmological constant...)
would be considerably modified.
\vskip 4mm
Gravitational coupling of superluminal particles to ordinary ones is 
expected to be
weak. Assuming that each superluminal sector
has its own Lorentz metric
$g_{[i] \mu \nu }$
($[i]$ for the $i$-th sector), with $c_i$ setting the speed scale, 
we may expect each sector to generate
its own gravity with a coupling constant $\kappa _i$ and a new 
sectorial graviton.
Gravitation would be a single and universal interaction in the 
limit where all $c_i$ tend to $c$
and where a single metric can be used, but not in other situations.
As an ansatz, we can assume that the static gravitational coupling 
between two different sectors
is lowered by a factor proportional to a positive power of the 
ratio between the two critical speeds
(the smallest speed divided by the largest one). 
Static gravitational forces between ordinary matter 
and matter of the $i$-th
superluminal sector would then be proportional to a positive power of 
$c/c_i$ which can be a very small number.
Experiment seems to suggest that other known gauge interactions 
(e.g. electroweak, strong) also couple weakly to
superluminal matter and that, if such matter exists, it can be 
produced only at very high energies
(future generations of accelerators?).
Interactions between ordinary and superluminal matter will break 
ordinary Lorentz invariance.
This will happen basically at very high energy and very short 
distance, and is not incompatible with the success
of conventional tests of relativity.
\vskip 6mm 
{\bf 2. THE RELEVANCE OF COSMIC RAY STUDIES}
\vskip 5mm
Because of the weak coupling to ordinary particles,
concentrations of superluminal matter would not necessarily follow 
the same pattern as "ordinary" galaxies and clusters of
galaxies (made of ordinary matter), nor would they need to occur at 
the same places. Although we
expect correlations between the distributions of ordinary and 
superluminal matter at large scales, it may be difficult
to reasonably assume that the gravitational role of galactic halos 
be due to superluminal
particles.
Superluminal particles can be very abundant and even provide most 
of the dark matter at cosmic scale,
but they will be extremely difficult
to detect if they interact very weakly with ordinary matter. 
Similarly, astrophysical objects made of superluminal matter may
elude all conventional observational techniques
and be extremely difficult to find. 
Cosmic rays can then play a crucial role.
\vskip 4mm
{\bf a) Ordinary primaries.} $Annihilation$ of pairs of slow 
superluminal particles
$into$ $ordinary$ $particles$, releasing very high kinetic energies 
from the superluminal rest masses,
would yield a unique $cosmic$ $signature$ allowing cosmic ray 
detectors to search
for this new kind of matter in the present Universe. 
$Collisions$ of high-energy superluminal particles $with$
$extra$-$terrestrial$ $ordinary$ $matter$ may also yield high-energy 
ordinary cosmic rays.
High-energy superluminal particles can be produced from acceleration, 
decays, explosions... in astrophysical objects made of
superluminal matter. Pairs of slow superluminal particles can also 
annihilate into particles of another superluminal sector
with lower $c_i$ , converting most of the rest energies into a 
large amount of kinetic energy.
Superluminal particles moving at $v>c$ will release $"Cherenkov"$ 
$radiation$ $in$ $vacuum$, i.e.
 spontaneous emission of particles of a
lower critical speed $c_i$ (for $v>c_i$) including ordinary ones,
providing a new source of (superluminal or ordinary) 
high-energy cosmic rays.
\vskip 4mm 
{\bf b) Superluminal primaries.} High-energy superluminal particles 
can directly $reach$ $the$ $earth$ and undergo
collisions inside the atmosphere, producing many secondaries like
ordinary cosmic ray events. They can also interact with the rock or 
with water near some underground or underwater detector,
coming from the atmosphere or after having crossed the earth, 
and producing clear signatures.
Contrary to neutrinos, whose flux is strongly attenuated by the 
earth at
energies larger than $10^6$ $GeV$ , superluminal particles will in 
principle not be stopped by earth at these energies.
Even if the interaction probability is expected to be very
weak, the rate for superluminal cosmic ray events can be observable 
if we are surrounded by important concentrations of superluminal
matter.
\vskip 4mm
The possibility that superluminal matter exists, and that it plays 
nowadays an important role in our Universe,
should be kept in mind when adressing the two basic questions 
raised by the analysis of cosmic ray events:
a) the nature of the cosmic ray primary; b) the identification of 
the source of the cosmic ray. If the primary
is a superluminal particle, it will escape conventional criteria 
for particle identification
and most likely produce a specific signature. If the
source is superluminal, it can be located anywhere 
(and even be a free particle) and will not be at the same place as 
conventional sources
of ordinary cosmic rays. High-energy cosmic ray events 
originating form superluminal sources
will for the first time provide hints on the location of such sources.
The energy dependence of the events should also be taken into account. 
At very high energies,
the Greisen-Zatsepin-Kuzmin cut-off [3] does not in principle hold for 
cosmic ray events originating from superluminal matter:
this is obvious if the primaries are superluminal particles
that we expect to interact very weakly with the cosmic microwave 
background,
but is also true for ordinary primaries as we do not expect them to
be produced at the locations of ordinary sources and there is no upper 
bound to their energy around $100~EeV$.
To date, there is no well established interpretation [3 , 4] of the 
highest-energy cosmic ray events [5].
Primaries (ordinary or superluminal)
originating from superluminal particles are acceptable candidates 
and naturally escape several problems
faced by cosmic rays produced at ordinary sources.
\vskip 5mm
{\bf References}
\vskip 4mm
\noindent
[1] L. Gonzalez-Mestres, "Properties of a possible class of particles 
able to travel faster than light", Proceedings of the
Moriond Workshop on "Dark Matter in Cosmology, Clocks and Tests of 
Fundamental Laws", Villars (Switzerland), January 21 - 28 1995 ,
Ed. Fronti\`eres. Paper astro-ph/9505117 of electronic library.
\par   
\noindent
[2] L. Gonzalez-Mestres, "Cosmological implications of a possible 
class of particles able to travel faster than light",
Proceedings of the IV International Conference on Theoretical and 
Phenomenological Aspects of Underground Physics,
TAUP95 , Toledo September 1995~, Ed. Nuclear Physics Proceedings. 
Paper astro-ph/9601090 of electronic library.
\par 
\noindent
[3] K. Greisen, Phys. Rev. Lett. 16 , 748 (1966); G.T. Zatsepin and 
V. A. Kuzmin, Pisma Zh. Eksp. Teor. Fiz. 4 , 114 (1966).
\par
\noindent
[4] F. W. Stecker, Phys.~Rev.~180~,~1264~(1969);
F. Halzen, R.A. Vazquez, T. Stanev and V.P. Vankov, 
Astropart.~Phys.~3~,~151~(1995);
T.J. Weiler and T.W. Kephart, paper presented at the Dark Matter'96 
Conference, UCLA February 14-16 1996 ,
astro-ph/9605156~.
\par
\noindent
[5] S. Yoshida et al.,~Astropart.~Phys.~3~,~105 (1995);
D.J. Bird et al. Astrophys. J. 424~, 491 (1994) and 
Astrophys. J. 441~, 144 (1995);
M.A. Lawrence, R.J.O. Reid and A.A. Watson, J. Phys. G17 , 733 (1991);
B.N. Afanasiev et al. in Proceedings of the XXIV International 
Cosmic Ray Conference, Rome 1995 , Vol. 2 ,
page 756.
\end{document}